\documentclass[reprint, aps, prl, 10pt, showpacs, showkeys]{revtex4-1}

\usepackage{amsmath, amssymb}
\usepackage{natbib}
\usepackage{graphicx}

\newcommand{\kb}{k_\mathrm{B}}
\newcommand{\fsolv}{f_{\mathrm{C}\parallel}}
\newcommand{\Tc}{T_\mathrm{c}}
\newcommand{\xiparallelAS}{\xi_\parallel^{\left(\mathrm{AS}\right)}}
\newcommand{\xiparallelS}{\xi_\parallel^{\left(\mathrm{S}\right)}}

\newcommand{\Tw}{T_\mathrm{w}}
\newcommand{\mgn}{\mathfrak{m}}
\newcommand{\mgnb}{\mgn_0}
\newcommand{\Last}{L^\ast}
\newcommand{\Laast}{L^{\ast\ast}}
\newcommand{\citen}[1]{\romannumeral-`\x\setcitestyle{numbers,open={},close={}}\cite{#1}\romannumeral-`\x\setcitestyle{numbers,square}}

\begin{document}
\title{Lateral critical Casimir force in 2D Ising strip with inhomogeneous walls}
\date{\today}
\author{Piotr Nowakowski}
\email{pionow@is.mpg.de}
\affiliation{Max--Planck--Institut f\"ur Intelligente Systeme, Heisenbergstr.\ 3, 70569 Stuttgart, Germany}
\affiliation{Institut f\"ur Theoretische Physik IV,Universit\"at Stuttgart, Pfaffenwaldring 57, 70569 Stuttgart, Germany}
\author{Marek \surname{Napi\'orkowski}}
\affiliation{Institute of Theoretical Physics, Faculty of Physics, University of Warsaw, ul.\ Ho\.za 69, 00--681 Warszawa, Poland}
\begin{abstract}
We analyze the lateral critical Casimir force acting between two planar, chemically inhomogeneous walls confining an infinite 2D Ising strip of width $M$. The inhomogeneity of each of the walls has size $N_1$; they are shifted by the distance $L$ along the strip. Using the exact diagonalization of the transfer matrix, we calculate the lateral critical Casimir force and discuss its properties, in particular its scaling close to the 2D bulk critical point, as a function of temperature, surface magnetic field, and the geometric parameters $M$, $N_1$, $L$. We determine the magnetization profiles which display the formation of the bridge joining the inhomogeneities on the walls and establish the relation between the characteristic properties of the lateral Casimir force and magnetization morphologies. We check numerically that breaking of the bridge is related to the inflection point of the lateral force.
\end{abstract}
\maketitle

Critical Casimir forces acting between the walls confining fluctuating thermodynamic systems have been discussed in the literature for nearly four decades \cite{fisher78, krech92, brankov00, hertlein08,  gambassi09}. The initial studies were mainly devoted to systems confined by chemically homogeneous boundaries \cite{krech92, burkhardt95, hanke98} and, in case of a slit geometry (with one or both walls homogeneous), the Casimir forces are perpendicular to the walls \cite{evans94, nowakowski09, toldin13}. 
When both walls are geometrically or chemically structured, the critical Casimir force can have a non--zero component along them \cite{sprenger06, trondle10, bimonte14}, the so--called lateral critical Casimir force. The presence of such a force may lead to interesting phenomena, e.g., formation of patterns among colloidal particles near non--uniform substrates \cite{soyka08}. In this letter we discuss the lateral critical Casimir force in equilibrium in a 2D Ising strip bounded by chemically inhomogeneous walls. As a result of the universality principle, the properties under investigation are relevant to either a confined one--component fluid close to criticality or a binary mixture close to its demixing point \cite{pelissetto02}. 

\begin{figure}[t]%
\begin{center}
\includegraphics[width=\columnwidth]{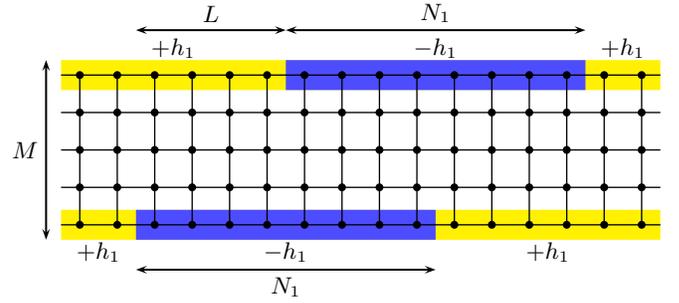}%
\end{center}
\caption{The inhomogeneous Ising strip with spins denoted by dots. The surface magnetic field modeling the influence of the walls, acts on the boundary spins only: the yellow (lighter) color denotes field $+h_1$ and the blue (darker) color denotes $-h_1$. The parameters $L$ and $N_1$ characterize the geometry of the inhomogeneities of the walls.}
\label{system}
\end{figure}

We consider the Ising strip consisting of $N$ columns and $M$ rows of a square lattice in which neighboring spins interact via ferromagnetic coupling constant $J>0$. There is no bulk magnetic field. The short--range interaction of the inhomogeneous confining walls with the ferromagnet is modeled by the surface magnetic field acting on spins in the first and the last row only. The surface field is piecewise constant on each wall: it is equal to $h_1>0$ except for a group of $N_1$ subsequent columns where it equals $-h_1$. There is only one inhomogeneity on each wall. The upper and lower wall inhomogeneities, i.e., the regions with inverted surface field, are shifted by distance $L$, see Fig.~\ref{system}. Periodic boundary conditions are imposed along the strip and the limit $N\to\infty$ is taken at fixed $M$ and $N_1$. The Hamiltonian has the form 
\begin{multline}
\label{Hamiltonian}
\mathcal{H}\left(\left\{ s_{m,n}\right\};h_1,M,L,N_1\right)=-J\sum_{m=1}^M 
\sum_{n=1}^N s_{m,n}s_{m,n+1}\\
-J\sum_{m=1}^{M-1}\sum_{n=1}^N s_{m,n}s_{m+1,n}-\sum_{n=1}^N\left(h_n
^{\prime}s_{1,n}+h_n^{\prime\prime}s_{M,n}\right),
\end{multline}
where the surface fields
\begin{equation}
h^\prime_n=\begin{cases}-h_1 & \text{for}\quad n=1,2,\ldots N_1,\\ +h_1 & 
\text{for}\quad n=N_1+1,N_1+2,\ldots,N,\end{cases}
\end{equation}
\begin{equation}
h^{\prime\prime}_n=\begin{cases}+h_1 & \text{for}\quad n=1,2,\ldots L,\\ -h_1 & 
\text{for}\quad n=L+1,L+2,\ldots,L+N_1,\\ +h_1 & \text{for} \quad 
n=L+N_1+1,L+N_1+2,\ldots,N.
\end{cases}
\end{equation}

To calculate the free energy $F\left(T, h_1, M, N, L, N_1\right)$ we apply the
method based on the exact diagonalization of the transfer matrix
\cite{kaufman49}, in which the surface fields are taken into account by adding
two additional rows of fixed spins \cite{abraham71, abraham73,
  maciolek96}. The change of sign of the surface fields, which corresponds to
the endpoints of the inhomogeneities of the walls, is generated by ``spin
flip'' operators additionally included in this approach \cite{abraham06}. The
resulting exact formulas for the lateral critical Casimir force depend on a
set of coefficients related to the transfer matrix method in the 2D Ising
strip. These coefficients have to be calculated numerically. This, in
principle, can be done with arbitrary precision. Therefore, many of our
results presented in this paper are numerical. Here, we skip all technical
details; they will be presented in Ref.~\citen{nn14}. The obtained formulas give the free energy of the system only for integer values of $L$. Thus, we define the dimensionless lateral Casimir force acting on the top wall via 
\begin{equation}
\fsolv\left(L\right)=-\frac{1}{\kb 
T}\lim_{N\to\infty}\left[F\left(L+\frac{1}{2}\right)-F\left(L-\frac{1}{2}\right)\right].  
\end{equation} 
This force is defined only for half--integer $L$. For simplicity, we do not
display here the remaining parameters $T$, $h_1$, $M$, $N$ and $N_1$, on which
the force depends. Note that negative values of $\fsolv$ imply that the force
is directed to the left in Fig.~\ref{system}. 

\begin{figure}[t]%
\includegraphics[width=\columnwidth]{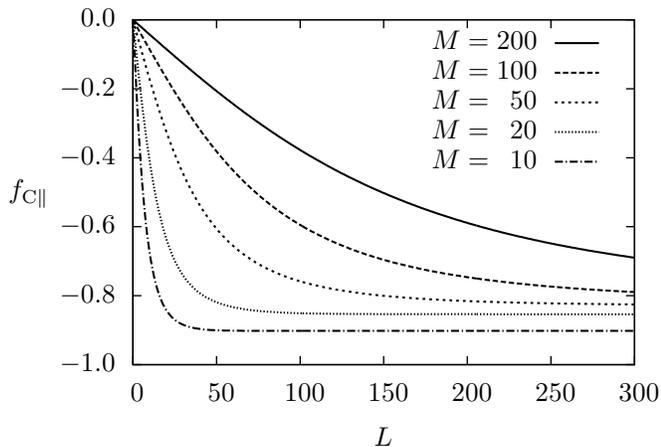}%
\caption{Plots of the lateral Casimir force for $N_1=\infty$, $T=0.8\,\Tc$ ($\Tc$ is the critical temperature of 2D square Ising model), $h_1=0.8\,J$ and different values of $M$. The force is defined for half--integer $L$ only, lines connecting points are plotted to guide the eye.}%
\label{fig2}%
\end{figure}

We first consider the case $N_1=\infty$. (Note that since the limit $N_1\to\infty$ is taken \textit{after} the thermodynamic limit, our analysis is restricted to $N_1\ll N$.) Typical plots of the force as a function of $L$ for different values of the strip width $M$ are shown in Fig.~\ref{fig2}. The lateral Casimir force is an odd function of shift $L$ and is negative for $L>0$. Upon increasing $L$ ($L>0$) the absolute value of the lateral force increases, and for $L\to\infty$ the force monotonously tends to $-2\sigma\left(T,h_1,M\right)$, where $\sigma(T,h_1,M)$ is the surface tension of a horizontal interface separating two phases present in a strip with homogeneous and opposite surface fields, $h^\prime_n = h_1 = - h^{\prime\prime}_n$ in \eqref{Hamiltonian} (antisymmetric case, AS) \cite{maciolek96}. This result is not surprising since in this limiting situation, two interfaces develop in the strip and they connect --- across the strip --- the endpoints of the inhomogeneities; for large $L \gg M$ each of them has approximately the length $L$. We have checked analytically, that the lateral force approaches its limiting value $-2\sigma\left( T, h_1, M\right)$ exponentially fast, i.e., $\fsolv\left(T, h_1, M, L, N_1\right)+2\sigma\left(T,h_1,M\right) \sim \exp\left[-L/\xiparallelAS\left(T,h_1,M\right)\right]$, where $\xiparallelAS\left(T,h_1,M\right)$ denotes the parallel correlation length in the antisymmetric case \cite{nn14,maciolek96}. 
 
\begin{figure}[t]%
\includegraphics[width=\columnwidth]{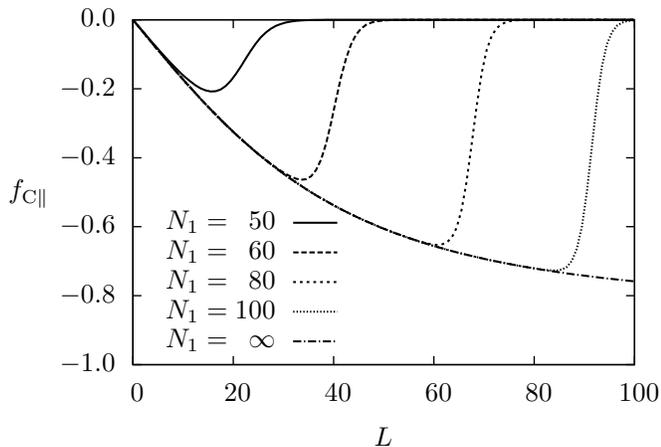}%
\caption{Plots of the lateral Casimir force for $T=0.8\,\Tc$, $h_1=0.8\,J$, $M=50$ and different values of $N_1$. Lines connecting points are plotted to guide the eye.}%
\label{fig5}%
\end{figure}

In the case of finite $N_1$, the behavior of the lateral Casimir force as function of $L$ is different. Fig.~\ref{fig5} presents plots of $\fsolv$ for different values of $N_1$. Upon increasing $L$, the absolute value of the force increases, reaches the maximum at (half--integer) $L=\Last\left(T,h_1,M,N_1\right)$, and then decreases to zero. The quantity $\Last<N_1$, and $\Last$ increases upon increasing $N_1$. For $L<\Last$ the force is almost independent of $N_1$ and practically the same as in the case of $N_1=\infty$, see Fig.~\ref{fig2}. For $L>\Last$ the force $\fsolv$ has an inflection point at $L=\Laast\left(T, h_1, M, N_1\right)$. This point is defined via inequalities 
\begin{equation}\label{Lastastdef}
\fsolv^{\prime\prime}\left(\Laast-\frac{1}{2}\right)>0>\fsolv^{
\prime\prime}\left(\Laast+\frac{1}{2}\right),\quad 
\Laast\in\mathbb{Z},
\end{equation}
and $\fsolv^{\prime\prime}\left(L\right)=\fsolv\left(L+1\right)-2\fsolv\left(L\right)+\fsolv\left(L-1\right)$ is the discrete version of the second derivative. We have checked analytically, that for $L> \Laast$ the lateral force displays asymptotic exponential decay $\fsolv\left(T,h_1, M, L, N_1\right)  \sim - \exp\left[- L/\xiparallelS\left(T,h_1,M\right)\right]$, where $\xiparallelS\left(T,h_1,M\right)$ is the correlation length in laterally homogeneous Ising strip with identical surface fields, $h^\prime_{n} = h_{1} = h^{\prime\prime}_{n}$ in \eqref{Hamiltonian} (symmetric case, S) \cite{nn14,maciolek96}. The decay observed for $L>\Laast$ becomes more abrupt upon increasing $N_1$ \cite{nn14}. The above described properties of the lateral Casimir force can be related to magnetization profiles in the strip; this will be discussed in the final part of this paper.

Now we turn to the analysis of the form of $\fsolv\left(T,h_1,M,L,N_1\right)$ in the scaling limit: $T\to\Tc$ and $M, L, N_1\to\infty$ with fixed values of $x=Mt/\xi_0^+$, $\lambda=L/M$, and $\eta_1=N_1/M$ \cite{evans94}. Here $t=(T-\Tc)/\Tc$ and $\xi_0^+=1/\left[2\ln\left(1+\sqrt{2}\right)\right]$ is the supercritical amplitude of the bulk correlation length, i.e., $x \sim M/\xi_\mathrm{b}\left(T\right)$. Note that in the scaling limit both $\xiparallelS$ and $\xiparallelAS$ are proportional to $M$ \cite{nn14,maciolek96}. We have checked numerically that in this scaling limit (in which the parameters $\lambda$ and $\eta_1$ can take real values) 
\begin{equation}\label{scalingf}
\fsolv\left(T,h_1,M,L, N_1\right)=\frac{1}{M}\ 
\mathcal{U}\left(x,\lambda,\eta_1\right)+\mathrm{O}\left(M^{-2}\right),
\end{equation} 
where the scaling function $\mathcal{U}$ does not depend on $h_1$ for $h_1> 0$. Similar independence of appropriate scaling functions of $h_1$ has been reported for laterally homogeneous Ising strips \cite{nowakowski09, abraham10}. Note that for large $M$ the lateral force decays as $1/M$, while the force perpendicular to the walls, evaluated per unit length of homogeneous slit, decays as $1/M^2$ \cite{evans94,nowakowski09}. 

\begin{figure}[t]%
\includegraphics[width=\columnwidth]{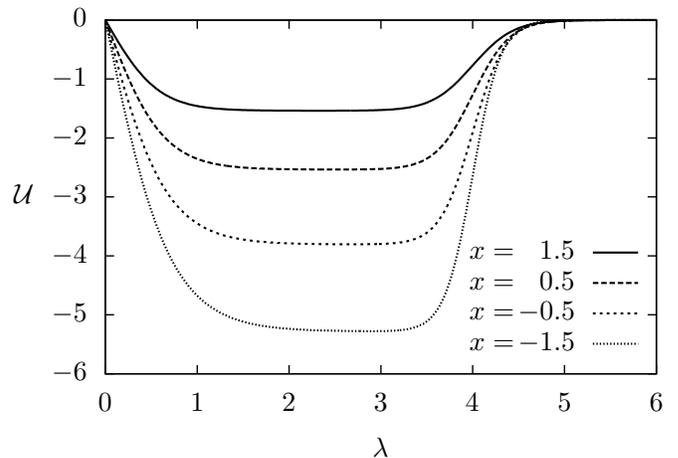}%
\caption{Plots of the scaling function $\mathcal{U}\left(x,\eta_1,\lambda\right)$ for $\eta_1=4.0$ and different values of $x$. Negative values of $x$ correspond to $T<\Tc$.}%
\label{figscale}
\end{figure}

By taking the limit $\eta_1\to\infty$ in \eqref{scalingf} we obtain the scaling form of the lateral force in the slit with infinite inhomogeneities (i.e., $N_1=\infty$ case)
\begin{equation}
\fsolv\left(T,h_1,M,L,\infty\right)=\frac{1}{M}\ 
\mathcal{U}\left(x,\lambda,\infty\right)+\mathrm{O}\left(M^{-2}\right).
\end{equation}
For large $\lambda$ the above force approaches $-2\sigma\left(T,h_1,M\right)$, and 
\begin{equation}
\lim_{\lambda\to\infty}\lim_{\eta_1\to\infty} 
\mathcal{U}\left(x,\lambda,\eta_1\right)=-2\mathcal{S}\left(x\right),
\end{equation}
where $\mathcal{S}\left(x\right)$ is the scaling function for the surface tension $\sigma\left(T,h_1,M\right)=\mathcal{S}\left(x\right)/M+\mathrm{O}\left(M^{-2}\right)$ \cite{maciolek96}. We stress the importance of the order of the limits in the above formula; for the opposite order one has $\lim_{\eta_1\to\infty}\lim_{\lambda\to\infty}\mathcal{U}\left(x,\lambda,\eta_1\right)=0$. A plot of the scaling function for $\eta_1=4$ is presented in Fig.~\ref{figscale}. Upon increasing $\lambda$, the scaling function first decreases, reaches plateau at $-2\mathcal{S}\left(x\right)$, and then, around $\lambda=\eta_1$ ($L=N_1$), starts to increase towards zero.

The behavior of $\fsolv$ as a function of $L$ can be related to the properties of magnetization profiles. For finite $N_1$ one observes two types of magnetization profiles, one relevant to small $L$, the other to large $L$. The first type of morphology corresponds to the presence of a capillary bridge of negative magnetization joining the inhomogeneities of the walls. The second type refers to two droplets of negative magnetization, each located near the inhomogeneity of the wall. The crossover between these two types is observed numerically for $L\approx N_1$. For the first type, the increase of $L$ induces the increase of the interface length and the force is of the order of $-2\sigma\left(T,h_1,M\right)$. For the second type, the free energy is almost independent of $L$ and the lateral Casimir force is very small. Since in a semi--two--dimensional system with short--range interactions there are no sharp transitions, the presented picture suggests that $\Last$ can be considered as a candidate for the crossover point. Below, we calculate numerically magnetization profiles and check that the minimum of the force $\Last$ is in fact not related to the breaking of the capillary bridge. Instead, we prove numerically that the inflection point $\Laast$ is the proper indicator of this crossover.  
 
\begin{figure*}[t]
\includegraphics[width=\textwidth]{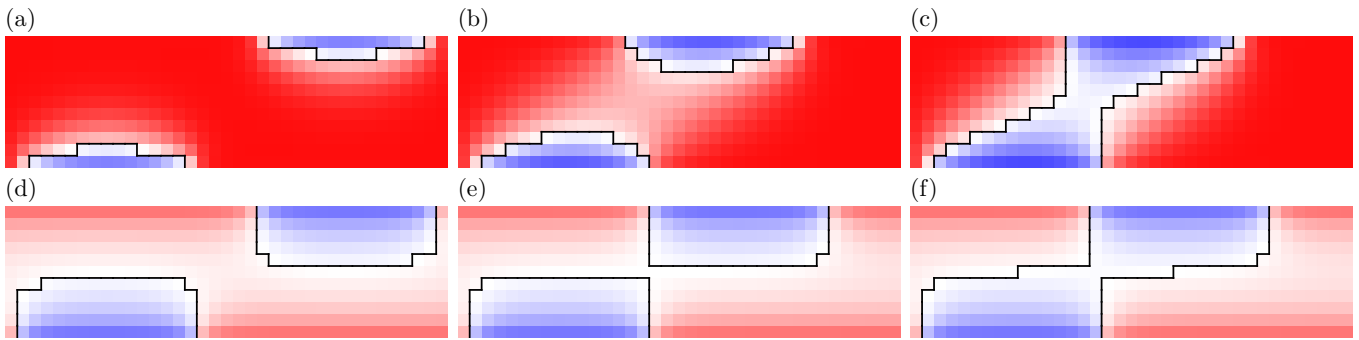}
\caption{Formation of a capillary bridge below (a)--(c) and above (d)--(f) the critical temperature $\Tc$ for different values of $L$. For all graphs $M=11$, $N_1=15$ and $h_1=0.8\,J$. (a) $T=0.8\,\Tc, L=20$, (b) $T=0.8\,\Tc, L=13$, (c) $T=0.8\,\Tc, L=12$, (d) $T=1.3\,\Tc, L=20$, (e) $T=1.3\,\Tc, L=15$, (f) $T=1.3\,\Tc, L=14$. The values of magnetization vary from $+1$ (red/dark) through $0$ (white/light) to $-1$ (blue/dark). Black lines denote interfaces.}%
\label{figmgn}%
\end{figure*}

The evaluation of magnetization profiles and analysis of their evolution under
the variation of the system parameters is done by transforming the appropriate
matrix elements to Pfaffians using the Wick's theorem~\cite{wick50}. The
Pfaffians are determined numerically by a specially developed method similar
to Refs.~\citen{stecki94} and \citen{gonzalez11}; the details will be
presented in Ref.~\citen{nn14}. The numerical complexity of the algorithm restricts the calculation of the magnetization profiles to strip widths below $M=26$. A typical scenario of the bridge formation is presented in Fig.~\ref{figmgn}. For a given magnetization profile we define the interface as the line that separates regions with opposite signs of magnetization \footnote{Note, that this definition of the interface becomes imprecise above $\Tc$. In the strip with opposite surface fields (AS) the interface is located exactly in the center of the system while magnetization remains practically zero everywhere except for the region near boundaries.}. It is marked by the black line in Fig.~\ref{figmgn}. 

Here, we report the results for $T>\Tw\left(h_1\right)$, the temperature of
critical wetting transition in the semi--infinite Ising model
\cite{abraham80,nowakowski09}. The case $T<\Tw$ will be discussed in Ref.~\citen{nn14}. 

For $\Tw\left(h_1\right)<T<\Tc$ and $L\gg N_1$ one observes two large droplets, one at each inhomogeneity of the walls. Upon decreasing $L$, the droplets first undergo deformations and then coalesce to from a bridge. However, for $T>\Tc$ one does not observe any substantial changes in the droplet shapes even when they are close to each other. The difference between these two scenarios can be explained by referring to the bulk magnetization of 2D Ising model $\mgnb\left(T\right)$. Above $\Tc$, $\mgnb=0$ and thus the magnetization in the region between droplets is close to $0$ while below $\Tc$, $\mgnb\neq0$ and the rapid change of magnetization in the region between undeformed droplets would increase the free energy; the droplets undergo deformations to minimize the effect.

\begin{figure}[t]%
\includegraphics[width=\columnwidth]{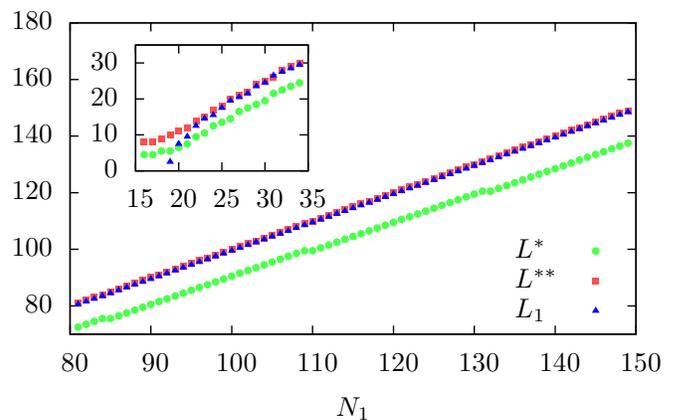}%
\caption{The dependence of the position of the maximum of the absolute value of the lateral Casimir force $\Last$, its inflection point $\Laast$, and the point of breaking of the capillary bridge $L_1$ as a function of the size of the inhomogeneities of the walls $N_1$; $T=0.8\,\Tc$, $h_1=0.8\,J$ and $M=21$. In this case, for $N_1\geq 25$ the difference between $\Laast$ and $L_1$ is $0.5$. For $N_1<19$ there is no capillary bridge and thus $L_1$ is undefined.}%
\label{fig7}%
\end{figure}

The point of breaking of the capillary bridge $L_1\left(T,h_1,M,N_1\right)$ is defined such that for $L=L_1-\frac{1}{2}$, there exists a region of negative magnetization connecting both walls, while for $L=L_1+\frac{1}{2}$ there is no such region; $L_1$ is always half--integer. Note that when $N_1<M$ there might be no bridge in the system, $L_1$ is in that case undefined, while the force still has the minimum and the inflection point. Fig.~\ref{fig7} displays $L_1$, $\Last$ and $\Laast$ as functions of $N_1$ for $T=0.8\,\Tc$, $h_1=0.8\,J$, and $M=21$. One observes that $\Last$ has a different asymptotic behavior than $L_1$ or $\Laast$: depending on the values of $T$, $h_1$, and $M$ the ratio $\lim_{N_1\to \infty}\Last/N_1$ can take any value between $0.5$ and $1$ \cite{nn14}. 

We have checked numerically \cite{nn14} that for $N_1$ large enough, $L_1=N_1-\frac{1}{2}$ and $\Laast=N_1$. This is the best possible agreement between these \textit{discrete} functions. The question is whether in the continuous model in the same universality class these functions would have the same asymptotic behavior. To answer this question, we define two new functions $\mathcal{M}_1$ and $\mathcal{M}^{\ast\ast}$ which are measures of the difference between $L_1$ and $\Laast$, and $N_1$,
\begin{align}
\mathcal{M}_1\left(N_1\right)&=\mgn\left(L=N_1,m=\frac{M-1}{2},n=N_1\right),\\
\mathcal{M}^{\ast\ast}\left(N_1\right)&=\fsolv^{\prime\prime}\left(L=N_1-\frac{1
}{2}\right)+\fsolv^{\prime\prime}\left(L=N_1+\frac{1}{2}\right),
\end{align}
where $\mgn$ denotes the magnetization in the slit. For simplicity, we do not display their dependence on $T$, $h_1$, $M$;  $M$ is assumed to be odd. The value of $\mathcal{M}_1$ is equal to the magnetization in the center of the system for $L=N_1$. This magnetization is positive when there is no bridge in the system and negative when there is a bridge. If one assumes that the lateral Casimir force is linear around $L=N_1$, then $\mathcal{M}^{\ast\ast}$ approaches $0$ as the inflection point approaches $N_1$. 

We have checked numerically that in the limit $N_1\to\infty$ with $T$, $h_1$ and (odd) $M$ fixed, both quantities $\mathcal{M}_1\left(N_1\right)$ and $\mathcal{M}^{\ast\ast}\left(N_1
\right)$ are proportional to $\exp\left(-N_1/\xiparallelAS\right)$. Thus we conjecture that in a continuous model the asymptotic behaviors of $L_1$ and $\Laast$ will be the same. The conjectured relation between the asymptotic behaviors of $L_1$ and $\Laast$ is useful because the numerical calculation of the critical Casimir force is much faster than the calculation of the magnetization profiles.

In a two--dimensional Ising strip confined by inhomogeneous walls, the component of the critical Casimir force parallel to the walls displays a non--monotonous behavior as function of the shift between the inhomogeneities of the walls $L$. We showed numerically that near the bulk critical point this lateral force takes the scaling form. We also studied the relation between the properties of the lateral force and the morphological transition taking place in the slit, i.e., the formation of the bridge joining the inhomogeneities of the walls.  For this purpose we calculated the magnetization profiles and compared the two values of $L$: $\Last$ for which the absolute value of the lateral force has a maximum and $\Laast$ for which this force has an inflection point, with the value $L_1$ for which the capillary bridge is formed. We checked numerically that $\Laast$ and $L_1$ display the same asymptotic behavior as functions of the inhomogeneities of the walls size, while $\Last$ behaves differently. This leads us to the conjecture that the inflection point of the lateral force can serve as the indicator of the morphological transition, and that --- upon increasing $L$ --- the absolute value of the lateral force starts to decay before the breaking of the capillary bridge takes place.

We thank Siegfried Dietrich and Matthias Tr\"ondle for helpful discussions and hospitality. Support from the (Polish) National Science Center via 2011/03/B/ST3/02638 is also gratefully acknowledged.


%

\end{document}